\documentclass[10pt]{article}
\usepackage{amsmath, amssymb, graphics, setspace}
\title{Self-gravity Correction to the Chandrasekhar Limiting Mass of White Dwarfs}
\author{{\it Anarya Ray}\\Department of Physics \\Presidency University, Kolkata, India\\ \\ {\it Pronobesh Maity}\\Department of Physics\\Presidency University, Kolkata, India\\ and \\ International Centre for Theoretical Sciences, Bengaluru, India \\ \\{\it Parthasarathi Majumdar}\\Department of Physics\\Ramakrishna Mission Vivekananda University\\Belur, West Bengal, India}

\begin{document}
\maketitle
\begin{abstract}

While computing the Fermi degeneracy pressure of electrons in a white dwarf star within the framework of hydrostatic equilibrium, we depart from the extant practice of treating the electrons as a free fermion gas, by including the effect of the star's self-gravity as an effective gravitational potential. By the star's self gravity, we mean the gravitational field due to the star itself, resulting from the mass of its constituent atoms, the mass of the atom being effectively the mass of the proton. Modifying the free particle Hamiltonian with this effective potential, we employ first order quantum mechanical perturbation theory to compute the degeneracy pressure, in order to study the effect of inclusion of this self-gravity of the star on the Chandrasekhar limiting mass.  The final effect is found to be non-trivial, but perhaps a shade too small to alter any major observational result. 
\end{abstract}

\section*{Introduction}

In his derivation of the limiting mass of a white dwarf star using hydrostatic equilibrium, Chandrasekhar\cite{ch1}-\cite{ch2} assumed that the electrons inside the star can be modelled as {\it free} quantum mechanical particles inside a box of length equal to the radius of the star. The electrons are of course assumed to be trapped in an infinite well, since, to escape outside from the boundary, the electrons would need to overcome a huge Coulombic potential. Such a potential is clearly absent inside, as the star as a whole is electrically neutral. But inside the box, unlike the electrostatic interaction, gravity does {\it not} cancel out. The electrons inside the well are not free but are actually residing in a {\it background gravitational potential}. Since this background potential is admittedly weak, the electrons can be more realistically modelled approximately as free particles inside a box, but {\it with an additional perturbation} by this background gravitational potential. The aim of this article is to explore the effects of such a background potential on the mass limit.  We argue that a correction to this limiting mass, albeit a small one, does indeed emerge within our proposed modification of electron dynamics inside the star.

We begin with a brief overview on white dwarfs and the Chandrasekhar mass limit, presenting a somewhat simpler but consistent derivation of the mass limit. Next we introduce the proposed modification due ot self-gravity of star, to the free electron Hamiltonian, within a sort of mean field picture. This modification is seen to be a weak effective gravitatonal potential. Resorting to first order time-independent perturbation theory, the effect of this effecive potential on the limiting mass of the star is then estimated, following the steps of Chandrasekhar's original derivation.

\section{Brief Overview of White Dwarfs and the Chandrasekhar Limit}

White Dwarfs have been a phenomenon of great importance to both theoretical and astrophysicists since their discovery in 1910. A White Dwarf is a stellar remnant, the fate of certain stars after they have exhausted all of their nuclear fuel.  They are extremely dense with masses comparable to that of the Sun but a volume comparable to that of the earth. Sir Arthur Eddington, a leading astrophysicist of his time, formulated the first theoretical challenge that came from the existence and stability of White Dwarfs, commonly known as "Eddington's paradox" which can be stated in the simple and clever words: "A star needs energy to cool".

The meaning of this statement is as follows. When a star runs out of nuclear fuel, it starts to collapse under its own gravity. Since there is no source of energy left, the radiating star looses energy and its temperature falls. As temperature falls, we expect the ionized stellar material to recombine and form normal atoms. But in order to form such atoms the star must expand to the density of normal atoms at that temperature, working against gravitational potential energy. In the words of Eddington, `` When the star cools down and regains the normal density ordinarily associated with solids, it must expand and do work against gravity. {\it The star will need energy to cool.}"

But a white dwarf's stellar material would have radiated so much energy that it cannot expand to normal densities that are associated with atoms at such low temperatures.  Such a star cannot be stable since it has nothing to sustain it any longer. Thus, according to such a scenario, the star will continue to collapse unabatedly, and one is led to conclude that stable white dwarfs ought not to exist. Yet, the existence of stable white dwarfs is in no doubt since this has been confirmed observationally for decades. Herein lies the paradox.  

This paradox was resolved by R.H. Fowler in his 1926 landmark paper entitled "Dense matter".  Fowler emphasized that as the temperature of the star falls, the electrons inside will become degenerate and by virtue of the Pauli Exclusion Principle will have a zero point energy and a corresponding degeneracy pressure which would prevent further collapse. 
 
In the following section, a somewhat simpler derivation of Fowler's results is worked out.  The conclusions were known since 1926 and are given here just for the sake of completeness.

\subsection{Fowler's Results}

In a completely degenerate electron gas, the density of states is calculated in the following way:  Since the electrons cannot escape the volume of the star (due to Coulomb interaction), they are assumed to be trapped in an infinite square well inside which they are free (since the star as a whole is electrically neutral,
the average electrostatic interaction cancels out inside the star). Under these circumstances, electrons in an infinite square well potential have the following energy
\begin{equation}
 E(n)=\frac{n^{2}\pi^{2}\hbar^{2}}{2ma^{2}} ~. \label{ener}
\end{equation}
where,
\begin{equation}
 n^{2}=n_{x}^{2}+n_{y}^{2}+n_{z}^{2} \nonumber 
\end{equation} 
and $n$ characterizes the energy level. To calculate the density of states, we imagine a sphere in the space of integers labeled by $n$; since $n_{x}, n_{y}, n_{z}$ are positive, and an allowed state occupies a unit cube in the first quadrant of the sphere in $n$-space. The number of states between energy level $n$ and $n+dn$ is:
\begin{equation}
g(n)dn= \frac{\pi n^{2} }{2}dn ~\label{denst}
\end{equation}
Since the occupation number of electrons in a state with energy $E$ is the Fermi-Dirac function which becomes a step function for
temperature tending to 0. The number of electrons occupying the energy levels between $n$ and $n+dn$ becomes
\[ 
  \ g(n)=\begin{cases}
         \frac{\pi n^{2} }{2} \quad   for \quad n\leq n_{f}  \\
         0      \qquad for \  n>n_{f}  
          \end{cases} 
    \] 
where $n_{f}$ is the Fermi occupation number, with the Fermi energy $E_{f}$ being defined from $E_f = E(n_f)$, using (\ref{ener}). Since each state can contain only two electrons (opposite spin) by virtue of the Pauli exclusion principle, the total number of electrons is given by:
\begin{eqnarray}
N =\int_{0}^{n_{f}} 2g(n)dn = \frac{\pi}{3} n_{f}^3 ~\Rightarrow~ n_{f}=\left( \frac{3N}{\pi} \right)^{1/3} \label{occ} 
\end{eqnarray}

Now we can compute the total kinetic energy of the electrons
\begin{eqnarray}
E_{kin} = \int_{0}^{n_{f}} 2g(n)E(n)dn =\frac{\pi ^{3} \hbar^{2}}{10ma^{2}} \left( \frac{3N}{\pi } \right)^{5/3} ~\label{kine}
\end{eqnarray}
But the total number of electrons N is equal to the total number of nuclei, which in turn is equal to the mass of the star divided by the mass of one nucleus.
\begin{equation}
E_{kin}=\frac{\pi^{3}\hbar^{2}}{10ma^2} \left( \frac{3M}{\pi A_m } \right)^{5/3} 
\end{equation}
where $M$ is the mass of the star, $a$ is the radius of the white dwarf and $A_{m}$ is the average mass of one nucleus of constituent particle of the star.
Thus,
\begin{equation}
E_{kin}=C\frac{M^{\frac{5}{3}}}{a^{2}} ~{\rm where} ~ C=\frac{\pi^{3} \hbar^{2}}{10m} \left( \frac{3}{\pi A_m } \right)^{5/3} ~. \label{kinec}
\end{equation}
The total energy $E_{tot}$ of the star is the sum of the kinetic and the self gravitational potential energy of the star,
\begin{eqnarray}
E_{Self-pot}&=& - \frac{3GM^{2}}{5a} \nonumber \\
E_{tot} &=& C\frac{M^{5/3}}{a^{2}}-\frac{3GM^{2}}{5a}
\end{eqnarray}

Minimizing the total energy gives  us  the mass radius equilibrium relation for White Dwarfs : 
\begin{equation}
    \frac{dE_{tot}}{da}=0~\Rightarrow M = \left( \frac{10C}{3Ga} \right)^3
\end{equation}
where C is given by equation (\ref{kinec}). The conclusion drawn from Fowler's work resolves Eddington's paradox. Of course a star needs energy to cool. At the low temperatures available to the star after it had radiated away a substantial portion of its thermal energy arising from thermonclear fusion,  the electrons become degenerate. By virtue of the Pauli Exclusion principle this degenerate electron gas must have a `zero-point' energy, since all electrons cannot have vanishing energy in the ground state of the system. This energy leads to a equilibrium condition for White Dwarfs of all masses.  The radius of course varies inversely with the cube-root of the mass.

\section{Chandrasekhar Limit}

But as the mass increases, so does the density of the star, as its radius falls inversely with the cube-root of its mass. At very high densities, the electrons at the threshold energies will have a velocity close to the speed of light and hence become relativistic, invalidating equation (1). This scenario was worked out by Subrhamanium Chandrasekhar and a brief over view of this is now given. From Fowler's result it is clear that, the heavier the star is, the smaller and hence denser it is. The Fermi momentum - the threshhold momentum of the electrons - is an increasing function of density:
\begin{eqnarray} 
p_{f}=\frac{ \pi \hbar n_{f}}{a} = \left( \frac{ 4\pi^3 \hbar^3  \rho}{A_{m}} \right)^{1/3} ~, \label{fermom}
\end{eqnarray}
For the average White Dwarf, $ \rho \simeq 4.19 \times 10^{6}~gm/cc$. This corresponds to momentum of the order of 1.3mc. At such high momenta the electrons have to be treated relativistically. Chandrasekhar showed that in the relativistic case, the White Dwarfs have a limiting mass. For any White Dwarf heavier than that mass, the electron degeneracy will not be able to prevent the collapse. We present below a somewhat simpler order of magnitude derivation of the limiting mass than the original version.

Starting with the well-known formula for the relativistic energy-momentum relation for a free particle of mass $m$, 
\begin{eqnarray}
E=\sqrt{ p^{2}c^{2}+m^{2}c^{4}} ~, \label{rele}
\end{eqnarray}
Expanding this in powers of $mc/p$, we get:
\begin{eqnarray}
E \approx pc(1+O(0.8)) \simeq \frac{n \pi \hbar c}{a}
\end{eqnarray}
Using eqn. (\ref{denst}) for the density of states, we obtain
\begin{equation}
    E_{kin}=\int^{n_{f}}_{0} 2E_{n}g(n)dn = \frac{C_{1}\hbar c}{a} \left( \frac{M}{A_m} \right)^{4/3}~{\rm with}~ C_1=\frac{\pi^{2}}{4}(\frac{3}{\pi} )^{4/3} ~. \label{ekin}
\end{equation}
    
Thus the total energy of the star is :
\begin{equation}
    E_{tot}=\frac{C_{1}\hbar c}{a} \left( \frac{M}{ A_{m}} \right)^{4/3} - \frac{3GM^{2}}{5a}
\end{equation}

Minimizing this total energy, i.e., setting $(dE_{tot}/da)=0 \quad {\rm and} \quad d^{2}E_{tot}/{da^2} > 0$, we see that for {\it stable} equilibrium to exist, the radial dependence cancels out and a mass inequality emerges :
\begin{equation}
M < M_{limit}= (5 C_{1})^{3/2} \frac{( \frac{ \hbar c}{G} )^{3/2}}{A_{m}^{2}}
\end{equation}
This limiting mass, known as the Chandrasekhar Mass Limit, sets an upper bound for the mass of a white dwarf star. Any such star heavier than this mass cannot exist in nature as it will collapse without resistance and form much more compact objects like neutron stars or black holes.

In the calculation of his mass limit, Chandrasekhar, following Fowler, assumed that the electrons inside the star are free, except that they cannot escape the boundary of the star. In the next section, we argue against the practicality of this assumption and propose a correction to the limiting mass in light of a more realistic model for the electrons.

\section{Proposed Correction}

While calculating the electrons' average kinetic energy, Chandrasekhar assumed that the electrons are trapped in an infinite square well inside which they are free. But in reality, they are not. The average Coulombic interaction may cancel out due to the fact that the star as a whole is electrically neutral.  But an effective gravitational potential exists inside the square well which, despite being weak, can alter the quantum mechanical properties of the electrons and hence their zero point energy. We show that the order of magnitude correction to the limiting mass is in fact computable.

An electron inside a star at a distance $r$ from the radius experiences a gravitational field ($F$) only due to the matter contained in a Gaussian sphere of radius $r$. Assuming uniform density for simplicity, the gravitational flux across the surface of the star is proportional to the stellar mass $M = (4/3) \pi r^3 \rho$, 
\begin{equation}
\oint_{S} \vec{F} \cdot {\hat n} dS =-4\pi G~ \frac43 \pi r^3 ~\rho ~\label{gau}
\end{equation}
so that, the gravitational force at every point inside the star can be written as
\begin{eqnarray}
\vec{F}(\vec{r})=- \frac{4 \pi G \rho \vec{r}}{3}\ for \  r\leq  a ~, \label{fin}
\end{eqnarray}
while, for locations outside the star,
\begin{eqnarray}
\vec{F}(\vec{r})=- \frac{4 \pi G \rho a^{3}}{3r^{2}}\hat{r}\ for  \ r> a~. \label{fout}
\end{eqnarray}
This leads to an effective self-gravitational potential affecting the elecron gas,
\begin{eqnarray}
V(\vec{r})=- \int_{\infty}^{r} {F}(\vec{r}) \cdot d\vec{r} 
\end{eqnarray}
The potential energy $U(\vec {r} )$ is just $V(\vec{r})$ times the mass $m$ of the electron : $U(r)=(2 \pi m G \rho r^2/3) - 2 \pi G \rho m a^{2}$. Thus, in this scenario, the electrons are trapped in an infinite square well potential, with a weak harmonic potential inside the star. One can treat this weak internal self-gravity potential as a perturbation on the un-perturbed free electron dynamics inside the star. 

The first order perturbation correction to the energy spectrum can be easily computed: 
\begin{eqnarray}
E^{1}_{n} &=& \int d^3r \Psi_n^* U(r) \Psi_{n} ~\nonumber \\
&=& \frac{4 \pi G \rho}{3}(-\frac{4}{3} m a^{2}-\frac{m a^{2}}{4 n^{2} \pi^{2}}) ~. \label{en1}
\end{eqnarray}
where $ \Psi _{n}(r)=\sqrt{\frac{2}{a}}\sin \frac{n \pi r}{a} $  are the nth level eigenfunctions of the infinite square well.
    
    But $$\rho=\frac{M}{\frac{4}{3} \pi a^{3}}$$
    Hence
    
    \begin{eqnarray}
    E_{n}^{1}=\frac{GM m}{a}(-\frac{4}{3}-\frac{1}{4 \pi^{2} n^{2}})=\frac{\omega}{a}(-\frac{4}{3}-\frac{1}{4 \pi^{2} n^{2}}) 
    \end{eqnarray}
    Where $$\omega=Gm M$$
    Thus the actual energy spectrum of the electrons is not (12) but rather:
    \begin{equation}
    E(n)=\frac{n \pi \hbar c}{a} +E_{n}^{1}=\frac{n \pi \hbar c}{a} +\frac{\omega}{a}(-\frac{4}{3}-\frac{1}{4 \pi^{2} n^{2}}) \equiv An -B -\frac{D}{n^{2}} ~\label{cor}
    \end{equation}
\noindent where 
\begin{eqnarray}
A  \equiv \frac{\pi \hbar c}{a}~,~ B \equiv \frac{4 \omega}{3a}~,~D  \equiv  \frac{\omega}{4 a \pi^{2}}~. \label{abd} 
\end{eqnarray}           
The numerical order of magnitude values of $A, B$ and $D$ have been calculated from the observed density and radius of typical white dwarf stars: $\rho \approx 10^{6} ~gm/cm^3~,~ a \approx 7000~km $ yielding $A \approx 10^{-32}~,~B \approx 10^{-23}~,~D  \approx 10^{-25} $. Here, the relatively large values of the constants $B$ and $D$, compared to that of the unperturbed constant $A$ raise a question of validity of the perturbative result. One expects that the perturbative result would not dominate the zeroth order unperturbed result corresponding to the original scenario of the free electron gas; thus
\begin{eqnarray}
E(n) > 0 \Rightarrow An > B + \frac{D}{n^{2}} ~. \label{pert}
\end{eqnarray}  
Since the correction terms decrease with increasing $n$, there must be a minimum $n=n_0$ such that $E(n_0)=0$; any value of $n>n_0$ is acceptable as a perturbative correction. 

To find $n_0$, note that one has to solve a cubic equation; this is done by employing Cardano's method, yielding $n_0 \simeq (B/3A)\approx 10^{8}~i.e ~ An_0 \simeq B/3 $. This implies that our perturbative correction is valid only for electron states lying within the domain $3n_0 < n < n_f$, and this is consistent with our analysis in the ultrarelativistic limit. The density of states is, of course, still given by (\ref{denst}).  

Referring back to (\ref{ekin}), the total kinetic energy of the electrons can be computed as before, with only the $n=0$ lower limit of the integration being replaced by $n_0$; this gives us the result
\begin{eqnarray}
E_{kin} &=& \int_{n_{0}}^{n_{f}} \left(An -B-\frac{D}{n^{2}} \right) \pi n^{2} dn \nonumber \\
&=& \pi A\frac{n_{f}^{4}}{4}\left[1-\left( \frac{n_{0}}{n_{f}} \right)^{4} \right]-\pi B\frac{n_{f}^{3}}{3}\left[1-\left( \frac{n_{0}}{n_{f}} \right)^3 \right] 
         -\pi Dn_{f}(1-\frac{n_{0}}{n_{f}}) ~. \label{ekin2}
\end{eqnarray}
Now, $n_{f} \approx N^{\frac{1}{3}} \approx 10^{20}$, since, $N \approx 10^{60} $, it follows that $ \frac{n_{0}}{n_{f}} \approx 10^{-12}$, so that all powers of $n_{0}/{n_{f}}$ can be safely ignored. 

With these approximations, and substituting the expressions for the constants $A~,B,~,D$, the total kinetic energy can be written as
\begin{eqnarray}
        E_{kin}=C_{1}\frac{\hbar c}{aA_{m}^{\frac{4}{3}}}M^{\frac{4}{3}} -C_{2} Gm_{e}\frac{M^{\frac{4}{3}}}{aA_{m}^{\frac{1}{3}}}-C_{3}Gm_{e}\frac{M^{2}}{aA_{m}}
~. \label{ekinf}    
\end{eqnarray}
where, $ C_{1}$ is as defined in (16), $C_{2}=(\frac{3}{\pi})^{\frac{1}{3}} \frac{1}{4\pi }~,~ C_{3} = \frac{4}{3}$

The total energy with the perturbative correction can now be written as 
\begin{eqnarray}
        E_{tot} &=& E_{kin} - 3G\frac{M^{2}}{5a} \nonumber \\
        &=& \frac{M^{\frac{4}{3}}}{a} \left( C_{1} \frac{hc}{ A_{m}^{\frac{4}{3}}} -C_{2}\frac{Gm_{e}}{A_{m}^{\frac{1}{3}}}\right) -\frac{M^{2}}{a}\left( \frac{C_{3}Gm_{e}}{A_{m}}+\frac{3G}{5} \right) ~\label{etotf}
\end{eqnarray}
Minimizing this new expression for total energy, i.e., setting $dE_{tot}/{da} =0~$ and $d^{2}E_{tot}/{da^2} > 0 $, the dependence on the stellar radius $a$ cancels out as before, leaving a limiting mass: 
    \begin{eqnarray}
        M<M_{limit} \approx (5 C_{1})^{3/2} \frac{( \frac{ \hbar c}{G} )^{3/2}}{A_{m}^{2}} \left( 1-\frac{10}{3} \frac{m}{A_{m}}\right) ~. \label{maslim}
\end{eqnarray}
The expression of the first two factors with first term in the last factor in parenthesis is the original limiting mass a' la Chandrasekhar, while the second term is our leading correction arising from self-gravity of the star producing an effective gravitational potential inside the star. Clearly, this dimensionless correction term is a ratio of the mass of the electron to that of the proton, i.e., of the order of $10^{-4}$, and hence substantially smaller compared to the original contribution. This is as may have been expected, and in a sense justifies the neglect of the physical effect discussed here, in the incipient analysis. However, the effect is not so small as to be completely ignorable, especially if future observational studies require more precise results than what is available from the incipient analysis.

\section{Conclusion And Pending Issues}
From our calculations, we conclude that
\begin{itemize}
\item The effect of a background gravitational potential on the electrons inside a White Dwarf is physical and produces a change in the mass-limit, the change being of the order of $10^{-4}$.
\item This change will also affect the absolute luminosity of Type-Ia Supernovae as calculated from the Mass-Limit. Since Type-Ia Supernovae act as Standard candles, our correction might have a significant effect in the measured value of the cosmological parameters. In light of the second point, considering type 1-a supernovae to be thermonuclear explosions of super chandrasekhar mass white dwarfs, the total energy released in such an explosion and hence the luminosity can be thought to be approximately proportional to the mass of its progenitor times the speed of light squared. E.g., the luminosity $L= \alpha M_{limit}c^{2}$, when 
our correction term is incorporated, becomes: $L^{*} \approx L(1-0.0001)$. This will change the measured value of the luminosity distance by:
\begin{eqnarray}
d_{L}^{*}=\sqrt{\frac{L^{*}}{4 \pi Flux}} \approx d_{L} (1-0.0001)^{\frac{1}{2}} \approx d_{L} (1-0.00005) 
\end{eqnarray}
The resultant change is indeed small but the corresponding change in cosmological parameters might be significant enough, given the ever-increasing precision currently being achieved in measurement of these parameters. From this standpoint, there seems to be scope for further research in the area.
\item We have restricted ourselves to the simplest possible corrections to the celebrated result, based mainly on Chandrasekhar's Nobel lecture; one might wonder : are there others - based on more complicated models - which might produce corrections of similar magnitudes ? This is a very pertinent point which has not been addressed here. We plan to consider such refinements in future. 
\item Similarly, for denser white dwarfs, one might wonder whether general relativity ought to be used, with the Dirac equation in a spherically symmetric background (where it is separable) providing a more precise estimate of the correction. Certainly a very important topic to be taken up in the near future.
\item As far as observations directly related to the mass limit is concerned, our knowledge is scanty, except for one reported observation which concludes that the data reveals a white dwarf about twice the limiting mass \cite{HOW}. It is likely that rotation and magnetic fields will produce a heavier white dwarf. In any event, these aspects have not been studied in this paper.
\item Another related question is : can similar corrections arise in hydrostatic equilibrium applied to more compact astrophysical objects like neutron stars ? We hope to report on these issues in the near future.
\end{itemize}

\section{Acknowledgements}

We acknowledge interesting and useful discussions with Muktish Acharya, Amitava Banerjee, Ritaban Chatterjee and Suchetana Chatterjee. One of us (PM) acknowledges very interesting correspondence with Amitabha Sen.

\section{Appendix}

\subsection{Solution of equation (24) using \textbf{Cardano's Method}}
$$An^{3} -Bn^{2}-D=0$$
Let $n=m+\frac{B}{3A}$. Substituting, we get:
$$m^{3}-\frac{B^{2}}{3A^{2}} m +[-\frac{2B^{3}}{27A^{3}} -\frac{D}{A}]=0$$
Now, using $\frac{2B^{3}}{27A^{3}} \approx 10^{26} ~ and \quad \frac{D}{A} \approx 10^{7}$   we get :
$$m^{3}-\frac{B^{2}}{3A^{2}}+\frac{2B^{3}}{27A^{3}}=0$$
Thus
$$m^{3} +pm + q =0$$
where,  $p=-\frac{B^{2}}{3A^{2}}$  and  $q=-\frac{2B^{3}}{27A^{3}}$
Now, following \textbf{Cardano's method}, let  $m=u+v$
then, 
$$u^{3} +v^{3} +3uv(u+v) +p(u+v)+q=u^{3} +v^{3} +(3uv+p)(u+v)+q=0$$

Now, since arbitrarily many pairs (u,v) can satisfy  $u+v=m$, without loss of generality, we can impose another condition on u and v such that  $uv=-\frac{p}{3}$.
Thus, the u and v that satisfy both  $u+v=m$  and   $uv=-\frac{p}{3}$ are unique and they can be evaluated as follows.
$$u^{3}+v^{3} +q=0$$
Or,
$$ u^{3}-\frac{p^{3}}{27u^{3}}+q=0$$
$$ \implies (u^{3})^{2}+qu^{3}-\frac{p^{3}}{27}=0$$ 
which is a quadratic in $u^{3}$ and has the solution
$$u^{3}=\frac{-q\quad \pm \quad \sqrt{q^{2}+\frac{4p^{3}}{27}}}{2}$$
In order to avoid negative values of $n=(u+v)-3a$ since no. of states cannot be negative, we ignore one root of u:
$$u^{3}=\frac{B^{3}}{27A^{3}}+\frac{B^{3}}{A^{3}} \sqrt{
\frac{1}{27^{2}}-\frac{1}{27^{2}}} = \frac{B^{3}}{27 A^{3}}$$
thus
$$u = \frac{B}{3A}$$ 
and 
$$v=-\frac{p}{3u} = -\frac{B}{3A}$$
Thus, m=u+v=0 and  $n=m+\frac{B}{3A}$
Thus 
$$n=n_{0}= \frac{B}{3A}$$

\end{document}